\documentstyle[prl,aps,epsf]{revtex}
\begin{document}
\draft
\twocolumn[\hsize\textwidth\columnwidth\hsize\csname @twocolumnfalse\endcsname

\title{Vortex Plastic Motion in Twinned Superconductors}

\author{J.~Groth, C.~Reichhardt, C.~J.~Olson, Stuart Field, and Franco Nori}
\address{Department of Physics, The University of Michigan, Ann Arbor,
Michigan 48109-1120}

\date{\today}
\maketitle
\begin{abstract}
We present simulations, without electrodynamical assumptions,
of $B(x,y,H(t))$, $M(H(t))$, and $J_c(H(t))$,
in hard superconductors, for a variety of twin-boundary pinning potential
parameters, and for a range of values of the density and strength of the
pinning sites.
We numerically solve the overdamped equations of motion
of up to $10^4$ flux-gradient-driven
vortices which can be temporarily trapped at  $\sim 10^6$ pinning centers.
These simulations relate macroscopic measurements ({\it e.g.},
$M(H)$, ``flame'' shaped $B(x,y)$ profiles)
with the underlying microscopic pinning landscape and the plastic 
dynamics of individual vortices.
\end{abstract}
\vspace*{-0.07in}
\vspace*{-0.37in}
\vskip2pc]
\narrowtext  

One of the major unsolved problems in high-$T_c$ 
superconductivity is flux pinning \cite{reviews}. Not only is flux pinning
of intrinsic interest, understanding it is necessary
in order to develop technological applications.
Since vortices are pinned by interacting with
defects, a great deal of work has focused on naturally occurring
defects in high-$T_c$ superconductors \cite{reviews,blatter-rmp,tps-review}.
In particular, it has been found that twin boundaries (TBs)
[2-19] form easily in the high-$T_c$ compound YBa$_2$Cu$_3$O$_{7-x}$ (YBCO).
Thus TBs have been intensively studied by numerous experimental
techniques, including Bitter decoration
\cite{bitter-dolan},
torque magnetometry \cite{torque-gyorgy}
magnetization
\cite{magnet-oussena,magnet-lairson,zhukov},
and magneto-optical imaging
\cite{duran_nature,vlasov-prl,duran-comment-prl,welp-comment-prl,%
welp-physicaC}.
In spite of this intense and multi-pronged effort, there is still
controversy concerning the effect of TBs
(see, e.g., the conflicting statements of \cite{magnet-oussena,fleshler}).
In particular, the magneto-optical measurements
by Duran {\it et al.} \cite{duran_nature}
and Vlasko-Vlasov {\it et al.} \cite{vlasov-prl}
seem contradictory:
\cite{duran_nature}
finds that TBs act as {\it easy-flow\/} channels
for longitudinal vortex motion,
while \cite{vlasov-prl} finds them to be {\it barriers}.
Moreover, resistivity measurements by Kwok {et al.} \cite{kwok}
suggest that pinning is enhanced within the TB.
Recent results
\cite{duran-comment-prl,welp-comment-prl,welp-physicaC}
confirm that the effect of TBs on vortex motion has a nontrivial
dependence on several variables.  First, there is an
angular dependence, i.e., a twin may act either as an easy flow channel or a
barrier depending on the angle between the flux front and the TB.
Second, experiments
\cite{torque-gyorgy,magnet-oussena,welp-physicaC}
indicate that the effective pinning of TBs is temperature dependent.
A complete picture is still lacking due to the limitations of the
techniques used so far (e.g., magneto-optics imaging only probes
the regime of low fields, while resistivity explores only
the high temperature regime).

An alternative technique for investigating the
microscopic behavior of flux
in a hard superconductor is to use computer simulations
(see, e.g., \cite{simulations,ray94,2d-bean}, and references therein).
Here, we present simulations of the {\it spatio-temporal\/}
evolution of rigid flux-gradient-driven flux lines in a
twinned superconductor such as YBCO.
We first describe a model for vortex-vortex, vortex-pin, and
vortex-twin interactions.
We then investigate how these {\it flux-gradient-driven} flux lines
travel through the sample in the presence of 
{\it both\/} correlated (TB) and uncorrelated (point defect) pinning.
 From the microscopic dynamics of individual vortices,
we compute several macroscopic measurable quantities
(e.g., $B(x,y,H(t))$, $M(H(t))$, $J_c(H(t))$)
and relate them with pinning parameters:
the angle $\theta$ between the TB and the sample edge, and
the ratio of pinning strengths inside and outside the TB,
which is expected to vary with temperature $T$ \cite{blatter-rmp}.

Our simulation geometry is that of an infinite slab of superconductor in
a magnetic field applied parallel to both the slab surface and the
TBs.  Thus, demagnetization effects are unimportant.  We also 
treat the vortices as stiff, so that we need to model only 
a two-dimensional (2D) slice ($x$-$y$ plane) of the 3D slab.
One can think of our rigid flux lines as representing the 
``center of mass" positions of real somewhat-flexible vortices, 
and our pinning in the bulk as representing the average of the 
pinning along the length of the real vortex.  
For a flexible vortex, the 
bulk pinning can be comparable to the pinning in the TB even 
for a large sample.
Vortices enter the sample at points where the local energy---as
determined by the local pinning and vortex interaction---is low.
We correctly model the vortex-vortex force interaction by using the
modified Bessel function $K_1(r/\lambda)$ with cutoffs at $0.1\lambda$
and $6\lambda$ ($\lambda$ is the penetration depth).
Thus, the force (per unit length)
on vortex $i$ due to other vortices (ignoring cutoffs) is
$
\ {\bf f}_{i}^{vv} =  
\ \sum_{j=1}^{N_{v}} \ f_0 \ K_{1}(
|{\bf r}_{i} - {\bf r}_{j}| / \lambda  ) \
{\bf {\hat r}}_{ij} \, .
$
Here, the ${\bf r}_{j} $ are the positions of the $N_v$ vortices within a
radius $6\lambda$,
$ \ {\bf {\hat r}}_{ij} =
({\bf r}_{i} - {\bf r}_{j}) / |{\bf r}_{i} - {\bf r}_{j}| $,
and
$
\ f_0 = \Phi_0^2 / 8 \pi^2 \lambda^3
$
\cite{reviews}.
Forces are measured in units of $f_0$, lengths in units of 
$\lambda$, and fields in units of $\Phi_0/\lambda^2$.
Notice that our simulation correctly models the driving force 
as a result of {\it local\/} interactions; no externally imposed 
``uniform" current is applied to the vortices.
Further details of the simulation are presented in \cite{2d-bean}.

{\it Uncorrelated Pinning.---}
The most important imperfections in YBCO single crystals are
random microscopic defects such as oxygen vacancies in the CuO$_2$ planes.
In order to model such a large number of defects,
we divide our system into a very fine (e.g., $1000 \times 1000$) grid,
where each grid site represents a pinning site.
The density of pinning sites, $n_p$, is between
$150/ \lambda^{2}$ and $496/ \lambda^{2}$ for all the runs
presented here. These densities are within experimentally determined values
\cite{bitter-dolan}.
The maximum pinning force $f^{\rm thr}_{l,m}$ on each grid site
$(l,m)$ is randomly chosen with a uniform probability distribution,
from $[0,f_p]$, where $f_p$ is the maximum possible pinning threshold force
(or strength).
If the site is in a TB, then the maximum possible pinning threshold force
is $f^{\rm TB}_p$.
When the net force acting on a vortex $i$ located on grid site
$(l,m)$ is less than the threshold force
$f^{\rm thr}_{l,m}$, then the vortex remains pinned in that grid site.
However, if the force acting on that vortex becomes larger
than the threshold force $f^{\rm thr}_{l,m}$, then the effective
pinning force drops to zero and the vortex moves continuously until it
encounters a pinning site that has a threshold force greater than
the net force acting on the vortex.
Thus the pinning force on each grid site acts like a stick-slip friction force
$
{\bf f}^{vp}_{i} =  - \ {\bf f}^{\rm net}_{i}
\ \Theta ( f^{\rm thr}_{l,m} - f^{\rm net}_i  ),
$
where ${\bf f}^{\rm net}_{i}$ is the net force on the vortex
(due to other vortices and possibly a TB)  and
$\Theta $ is the Heaviside step function.  In other words,
$ {\bf f}^{vp}_{i} =  - \ {\bf f}^{\rm net}_{i}$ \ if
$ f^{\rm net}_i < f^{\rm thr}_{l,m} $, and
$ {\bf f}^{vp}_{i} =  0 $ if
$ f^{\rm net}_i \geq f^{\rm thr}_{l,m} $.

{\it Twin Boundary Pinning.---}
We have tested a very large variety of models for TBs, and
compared simulation results to experiment.
For the sake of concreteness, we focus here on the model which was most
consistent with experimental data; the results of the other models
will be discussed elsewhere. 
In this model, a TB consists of an attractive well containing
pinning sites that have a different maximum threshold force (or strength)
$f_p^{\rm TB}$ than that for sites outside the TB, $f_p$.
The ratio $f_p^{\rm TB}/f_p$
is expected to vary as a function of temperature $T$
because of the reduced dimensionality of the thermal fluctuations
for vortices in a TB \cite{blatter-rmp}.
The ratio is small (i.e., pinning in the twin is weak)
for low $T$, and it is larger at higher $T$ 
(see, e.g., page 1321 of \cite{blatter-rmp}).
By varying this ratio, we can mimic some of the
effects due to temperature.

For all the results presented here, we model the attractive TB
well as a parabolic channel with a width $2 \xi^{\rm TB}$.
The attractive force on the $i$th vortex due to the well of
the $k$th twin is thus
$
{\bf f}^{vTB}_{i} = f^{\rm TB}( d_{ik}^{\rm TB}/{\xi^{\rm TB}} ) \
\Theta ( ({\xi^{\rm TB} -       d_{ik}^{\rm TB})/\lambda} ) \;
{{\bf {\hat{r}}}_{ik}} \, ,
$
where $d_{ik}^{\rm TB}$ is the perpendicular distance between
the $i$th vortex and the $k$th TB, and $f^{\rm TB}$ is the
maximum force that the well of the twin exerts on a vortex
(i.e., the force needed to escape the TB).
The pinning inside the twin channel is modeled identically to the
pinning in the rest of the sample, except that it has a different
maximum strength, $f_p^{\rm TB}$ instead of $f_p$.
This microscopic model of TB pinning is very similar to the one
inferred from recent measurements in \cite{magnet-oussena},
where it is found that the TB ``channel'' has strong ``depth'' variations.

{\it Dynamics.---}
The overall equation for the overdamped motion of a vortex moving with a
velocity {\bf v} and subject to vortex-vortex and vortex-pinning forces
(due to intrinsic pinning $f_i^{vp}$ and TB pinning $f_i^{vTB}$) is
$
\ {\bf f}_{i} = {\bf f}_{i}^{vv} + {\bf f}_{i}^{vp}  + {\bf f}^{vTB}_{i}
= \eta {\bf v}_{i}\; ,
$
where
$
 {\bf f}_{i}   =
\ ({\bf f}^{vv}_{i} + {\bf f}^{vTB}_{i} )
\ [ 1 - \  \Theta ( f^{\rm thr}_{l,m} - f^{vTB}_{i} - f^{vv}_i ) ],
$
and we take $\eta = 1$.

Many parameters can be varied, making the systematic study of this
problem quite complex.  Here we chose to vary only the two most critical
variables, the angle between the TB and the flux flow, as  stressed in
\cite{duran-comment-prl,welp-comment-prl,welp-physicaC},
and the ratio $f_p^{\rm TB}/f_p$.
We find the critical current, $J_c$, for an untwinned sample
with $f_p = 0.2 f_0$ and $\lambda=1400$\AA,
to be $1.26 \times 10^6$ A/{cm}$^2$, which is
reasonable for YBCO at low $T$.
Thus we varied $f_p$ between $0.15 f_0 - 0.4 f_0$.
A more detailed investigation with different
TB widths, TB depths, intrinsic pinning strengths and different TB
potentials will be presented elsewhere. 
For the runs presented here, we fixed the depth of the TB channel
by setting the maximum TB force $f^{TB}$ to $0.6 f_0$, and the width of
the channel $2 \xi^{TB}$ to be approximately $0.7 \lambda$.

{\it Sample.---} We begin by studying in detail
how vortices entering a sample interact with a single TB.
The actual sample region is heavily pinned, and covers the region
with $(x,y)$ coordinates $0 \leq x \leq 66 \lambda $ and
$0 \leq y \leq 55 \lambda $ (Fig.~1).
The number of pinning grid sites in the sample is
$\frac{5}{6} \times 10^6 = 833,333$.
The linear size of each pinning grid site is
$d_p = 66 \lambda / 1000= 0.066\lambda$,
and the pinning density is $n_p = 1/d_p^2 = 230/\lambda^2$ which,
for a computer simulation, is extremely large.
Fig.~4 shows results for other samples.

{\it Flux Density Profiles and Current Flow.---}
Figure 1 shows the computed flux distribution for $f_p = 0.15 f_0$ with
$f_p^{\rm TB}/f_p = 3/4$ (a,b,c) and $f_p^{\rm TB}/f_p = 1/4$ (d).
The TB is at an angle $\theta = 60^{\circ}$ measured from the horizontal
sample edge located at the bottom, where flux enters
(note the similarity to the experimental situation in \cite{vlasov-prl}).
Fig.~1(a) shows the positions of the vortices after the
external field has been ramped up from zero.
Two important features stand out.  First, we see
that the vortices ``pile up'' on the side of the twin where
the flux enters. Second, on the other side of the TB we see
the ``shadow'' effect; the density of vortices is lowered.
These features of the vortex plastic flow are more explicit in 
(b), which shows the induction $B$ along the horizontal line in (a).
When the external field is ramped down to zero we also observe some
shadowing and pile up effects, but they are weaker since the number
of vortices in the sample is now smaller.
Both of these effects for the ramp up and down cases
are clearly observed in the experiments of \cite{vlasov-prl}.
Since we know the exact position of the vortices, we can calculate
the $B$ field due to the vortices, and hence the
{\it current flow\/} $J(x,y,H(t))$
(contours of constant $B$) near the TB. Fig.~1(c) shows the current flow
for the TB in (a), while (d), corresponding to lower $T$,
presents a ``flame''-shaped current flow similar to the ones
observed in several experiments
\cite{duran_nature,vlasov-prl,duran-comment-prl,%
welp-comment-prl,welp-physicaC}.

In Fig.~2 we show the positions
of the vortices after the field has been ramped up from zero for several
cases. In the top panels, the ratio $f_p^{\rm TB}/f_p$ is held
fixed, and the angle $\theta$ is varied; 
we see that the $\theta = 20^{\circ}$ TB is 
a barrier to flux flow, while the other angles are easy-flow
channels. Our model, like experiments
(e.g., \cite{welp-comment-prl,welp-physicaC}),
exhibits TB vortex flow which has an angular dependence.
In the lower panels of Fig.~2 the ratio $f_p^{\rm TB}/f_p$ is
increased. We still observe an angular dependence, but the
angle at which a twin becomes an easy-flow channel has increased.
As the ratio $f_p^{\rm TB}/f_p$ increases, the angle
$\alpha$ between the TB and the
direction of vortex flow must decrease if the twin is
to be an easy flow channel.

Figure 3 summarizes results obtained from many runs.
Fig.~3(a) presents the ``vortex plastic motion''
phase diagram
showing the region where the TB is an easy-flow channel (barrier)
for longitudinal vortex motion, indicated by open circles (solid squares).
The boundary, indicated by grey diamonds, corresponds to slow vortex motion.
This phase diagram quantitatively summarizes how
$\theta$ and $f_p^{\rm TB}/f_p$ affect the vortex motion.
As pinning in the twin increases, we expect the flux-front velocity
to decrease.  This observation is quantified in Fig.~3(b).  There,
$v^{\rm TB}$ ($v$) is the velocity of the flux front inside (outside) the
TB, measured along the direction perpendicular to the sample edge.
Figure 3(b) shows $v^{\rm TB}/v$ decreasing monotonically
for increasing $f_p^{\rm TB}/f_p$.
Monitoring the dynamics of individual vortices, we observe that the
longitudinal vortex transport in the TB is {\it plastic\/} 
(many vortices ``tear away" from the flux lattice when 
$v^{\rm TB}/v \neq 1$) and proceeds via two mechanisms:
(i) the actual motion of vortices along the TB, and
(ii) vortices from the bulk (and close to the TB vortex front)
falling into the TB.  Finally, our phase diagram [Fig.~3(a)]
suggests that the contradictory experimental results in 
\cite{duran_nature,vlasov-prl} can be understood as arising 
from probing different parameter regimes.

Refs.~\cite{welp-comment-prl,welp-physicaC} made the interesting observation
that flux enters the sample in a ``flame'' pattern around the TB.
\cite{welp-physicaC} suggests,
using current-conservation arguments, that this happens
whenever the twin acts as easy-flow channel.
We also observe flame patterns (see Fig.~4(a))
whenever the twin is an easy-flow channel.
The flame is caused by vortices {\it escaping}
from the TB, not gliding along it.

{\it Magnetization.---}
One clear advantage of our simulation is that we can obtain direct
{\it spatio-temporal\/} information on the distribution of
individual flux lines {\em inside} the sample. This is quite
difficult to obtain experimentally, especially for bulk samples.
What is typically measured in experiments is the average magnetization
$M$ over the sample volume.  In our simulation, we thus calculate
$ M = (4 \pi V)^{-1}  \int (H - B) \ dV$.
In Fig.~4(b) we show partial magnetization loops for
a sample where the TBs are perpendicular to the flux flow (d),
one where they are parallel to the flux flow (e),
and a sample without any twins at all.
Note that Fig.~4(d,e) are two particular (and extreme) cases
considered in the phase diagram Fig.~3(a):
$\theta = 0^{\circ}$ and $90^{\circ}$.
For these runs the sample is $42\lambda \times 42\lambda$,
$f_p^{\rm TB}/f_p = 0.1$, and $f_p =0.2 f_0$.

The loop where the twins are perpendicular to the flux flow
is clearly wider than the other two. This is because
a twin is a barrier to perpendicular vortex flow since
the vortices would have to cross the TB, which is a parabolic channel.
The loop where the TBs are parallel to the flux flow is narrower because
these TBs are easy flow channels.
Also, we repeated the above experiments at a higher ratio,
$f_p^{\rm TB}/f_p = 1.5$, and
found that the perpendicular TBs are still barriers to flux flow,
i.e., the magnetization is still higher than the untwinned
case. The parallel twins no longer act as easy-flow channels
and thus the magnetization is nearly identical to the
untwinned case.
All these results are qualitatively similar to those in
Ref.~\cite{magnet-oussena};
the minor differences are most likely due to finite size effects.

To summarize, we have performed simulations of flux-gradient-driven
vortices in twinned superconductors with a very large number of
($\sim 10^6$) point defects.  Experiments can be explained by 
using a simple model for a twin boundary: a parabolic well 
containing pinning sites which are different in strength from 
the rest of the sample.  In agreement with experiments,
we show that the twin may act as either a barrier or
an easy flow channel (producing ``flame''-type flux patterns),
depending on the angle between the flux flow and the
pinning ratio, $f_p^{\rm TB}/f_p$.
We quantify these results in detail and provide a phase diagram 
for the plastic motion of vortices [Fig.~3(a)].
When the twin is a barrier, it acts to guide flux motion
resulting in the pile up and shadowing effects.
We also compute magnetization loops for samples
where the twins are parallel and perpendicular to the flux flow and
obtain results similar to recent experiments \cite{magnet-oussena}.

SF was supported in part by the NSF under grant No.~DMR-92-22541.

\newpage

\begin{figure}
\centerline{
\epsfxsize=3.2in
\epsfbox{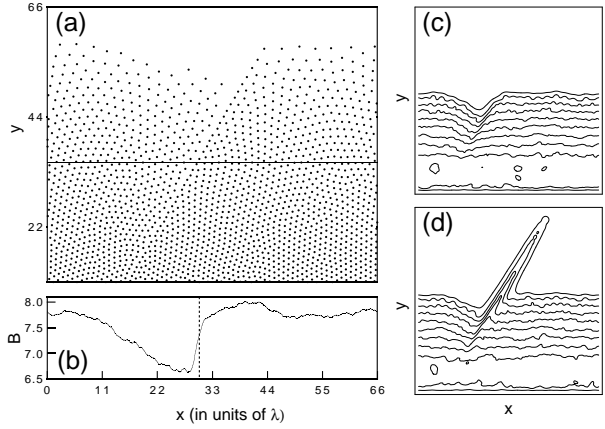}}
\vspace{0.05in}
\caption{
(a) Snapshot during the initial ramp-up phase of the
top view of the region where flux lines, indicated by dots, move.
(b) $B(x)$ along a $1.0 \lambda$--wide horizontal cut through the 
sample centered on the horizontal line $y=36\lambda$ indicated in (a).
We calculate $B(x,y)$ at each point due to all the vortices
and then find $B(x)$ by averaging over a strip, i.e.,
$B(x,y=y_0)= (\Delta y)^{-1} \int_{y_0}^{y_0+\Delta y} d\!y \, B(x,y)$.
This $B(x)$ clearly shows the ``shadowing'' and ``pile up'' effects
discussed in \protect\cite{vlasov-prl}.  
(c,d) Current flow (contours of constant B) around the
TB in (a) for several pinning strengths.
In (c) $f_p^{\rm TB}/f_p = 3/4$, as in (a), 
more effective source of pinning, 
and the TB is a barrier to longitudinal vortex motion.
In (d), $f_p^{\rm TB}/f_p = 1/4$ 
and the TB is an easy-flow vortex channel.
} \label{fig1}\end{figure}

\begin{figure}
\centerline{
\epsfxsize=2.3in
\epsfbox{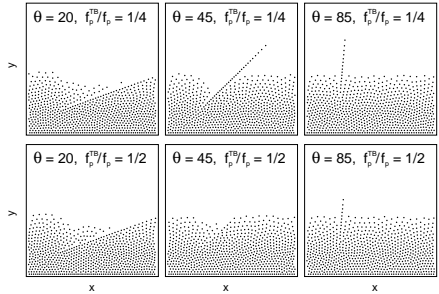}}
\vspace{0.05in}
\caption{
Top view of the region where flux lines, indicated by dots, move in the ramp-up
phase.  We have varied $\theta$ and $f_p^{\rm TB}/f_p$, and here we present
some of our results for
$\theta=20^\circ$ (left), $45^\circ$ (center), $85^\circ$ (right),
and $f_p^{\rm TB}/f_p=1/4$ (top), $1/2$ (bottom).
For the top panels the TB acts as an easy-flow channel,
except $\theta=20^\circ$, while for the bottom ones
{\it only} the sample with $\theta=85^\circ$ exhibits easy-flow
channel behavior. Here $f_p = 0.15 f_0$ and $n_p = 230/{\lambda}^2$.
} \label{fig2}\end{figure}

\begin{figure}
\centerline{
\epsfxsize=3.0in
\epsfbox{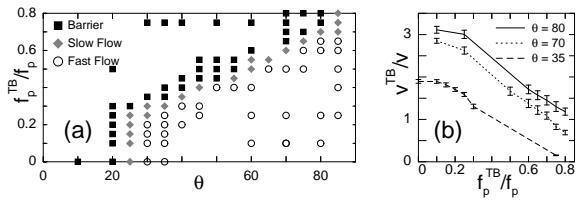}}
\vspace{0.05in}
\protect\caption{
(a) Phase diagram $f_p^{\rm TB}/f_p$ versus $\theta$ showing the region
where the TB is an easy-flow channel (barrier) for longitudinal vortex
motion, indicated by open circles (solid squares).
The boundary, indicated by grey diamonds, corresponds to slow vortex motion.
The ratio $v^{\rm TB}/v$
is $\approx 1$ (black squares),
between $1$ and $1.25$ (grey diamonds),
and larger than $1.25$ (open circles).
(b) $v^{\rm TB}/v$ versus $f_p^{\rm TB}/f_p$ for
$\theta=35^{\circ}, \; 70^{\circ}, 80^{\circ}$ (upper curve).
}\label{fig3}\end{figure}

\begin{figure}
\centerline{
\epsfxsize=3.0in
\epsfbox{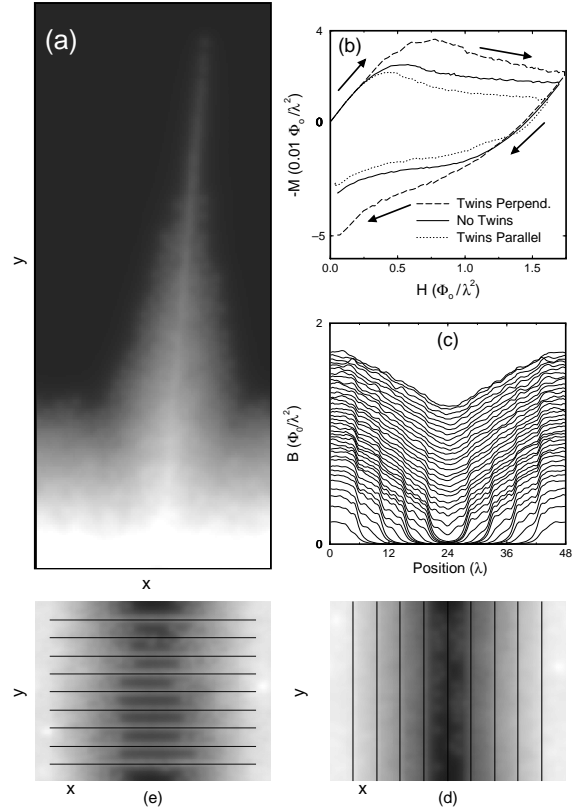}}
\vspace{0.15in}
\protect\caption{
(a)
$B_z(x,y)$ along a TB at $\theta = 85^{\circ}$ with 
$f_p^{\rm TB}/f_p = 0.05$, $f_p = 0.4f_0$, and 
$n_p = 152/{\lambda}^2$. $B_z(x,y)$ exhibits a ``flame''
front profile. The same-color contours of constant $B_z$ 
indicate the current flow around the twin.
The images 4(a,d,e) show $B_z(x,y)$, which is obtained by
approximating the field due to each vortex by a monopole
located $1\lambda$ below the surface and computing 
$B_z(x,y)$ at $z = 0.6\lambda$ above the surface
(see Ref.~\protect\cite{imaging-hess}).
(b)
$M(H(t))$ for heavily twinned samples; the TBs are 
$5.6 \lambda$ apart.  TBs, indicated by black lines in 
(d) and (e), act as barriers (for TBs perpendicular to 
the vortex flow as in (d)) or as easy-flow channels 
(for TBs parallel to the vortex flow as in (e)).
For comparison, $M(H(t))$ for an untwinned  sample is 
also shown in (b). 
In (b-e), $f_p^{\rm TB}/f_p = 0.1$, $f_p = 0.2f_0$,
and $n_p = 496/{\lambda}^2$.
(c)
$B(x)$ for the sample shown in (d).
Notice the {\it sudden decrease\/} in $B(x)$
(leading to high-currents) at each of the TBs.
This ``terraced'' $B(x)$ profile produces a periodic array of alternating
high and low current regions (see Ref.~\protect\cite{pinch-effect}).
In (c,d) the TBs act as a series of ``flux dams''.
}\label{fig4}\end{figure}

\end{document}